\documentclass{iau}
\usepackage{graphicx}

\title[IAU-S342.-- Molecular gas filamentary structures] 
{Molecular gas filamentary structures in galaxy clusters}

\author[F. Combes]   
{Francoise Combes$^1$}

\affiliation{$^1$Observatoire de Paris, LERMA, College de France, CNRS, PSL Univ., Sorbonne Univ.,
F-75014, Paris, France \\ email: {\tt francoise.combes@obspm.fr}}

\pubyear{2018}
\volume{342}  
\jname{Perseus in Sicily: from black hole to cluster outskirts}
\editors{K. Asada, E. de Gouveia dal Pino, H. Nagai, R. Nemmen, M. Giroletti, eds.}
\begin{document}

\maketitle

\begin{abstract}
Recent molecular line observations with ALMA and NOEMA in
several Brightest Cluster Galaxies (BCG) have revealed the
large-scale filamentary structure at the center of cool core clusters.
These filaments extend over 20-100kpc, they are tightly
correlated with ionized gas (H$\alpha$, [NII]) emission, and have
characteristic shapes: either radial and straight, or also
showing a U-turn, like a horse-shoe structure.
   The kinematics is quite regular and laminar, and the derived infall
time is much longer than the free-fall time. The filaments
extend up to the radius where the cooling time becomes larger
than the infall time. Filaments can be perturbed by the
sloshing of the BCG in its cluster, and spectacular cooling wakes
have been observed. Filaments tend to occur at the border
of cavities driven in the X-ray gas by the AGN radio jets.
Observations of cool core clusters support the thermal instability scenario,
which accounts for the multiphase medium in the upper atmospheres of BCG,
where the right balance between heating and cooling is reached,
and  a chaotic cold gas accretion occurs.
Molecular filaments are also seen associated to ram-pressure stripped
spiral galaxies in rich galaxy clusters, and in jet-induced star formation,
suggesting a very efficient molecular cloud formation even in hostile cluster environments.
\keywords{ galaxies: clusters: general --  galaxies: ISM -- galaxies: jets  -- galaxies: nuclei  --  galaxies: peculiar}
\end{abstract}


The Perseus cluster, the mascot of our Symposium, is indeed the prototype of cool core clusters,
where all physical phenomena can be studied in exquisite detail, due to its proximity.
Fabian et al. (2003)  have published a beautiful X-ray image, showing clearly the
cavities sculpted by the radio jets, and their unsharp-masked image revealed multiple ripples, 
tracing acoustic waves, transferring the AGN energy into the ICM (Intra-Cluster Medium), and
contributing to regulate the cooling.  Large quantities of molecular gas,  
gas, M(H$_2$) $\sim$ 10$^{10}$ M$_\odot$, have been detected
with the IRAM-30m telescope (Salom\'e et al. 2006), and there is a good
correlation between the CO and H$\alpha$ filaments, as if the ionized gas was
the interface between the dense cold gas and the ICM. Most of the H$\alpha$
is excited by shocks and cosmic rays, but there are also patches of
 star formation (Canning et al. 2014). The H$_2$
 1-0 S(1) ro-vibrational transition, tracing the shocked and highly excited molecular gas,
 has been mapped in the near-infrared (2 $\mu$m) with WIRCAM at the CFHT
 by Lim et al. (2012). The correspondence between these H$_2$ filaments and the
 H$\alpha$ ones over 85 kpc in length is striking.

\begin{figure}[h]
\begin{center}
  \includegraphics[width=0.4\textwidth]{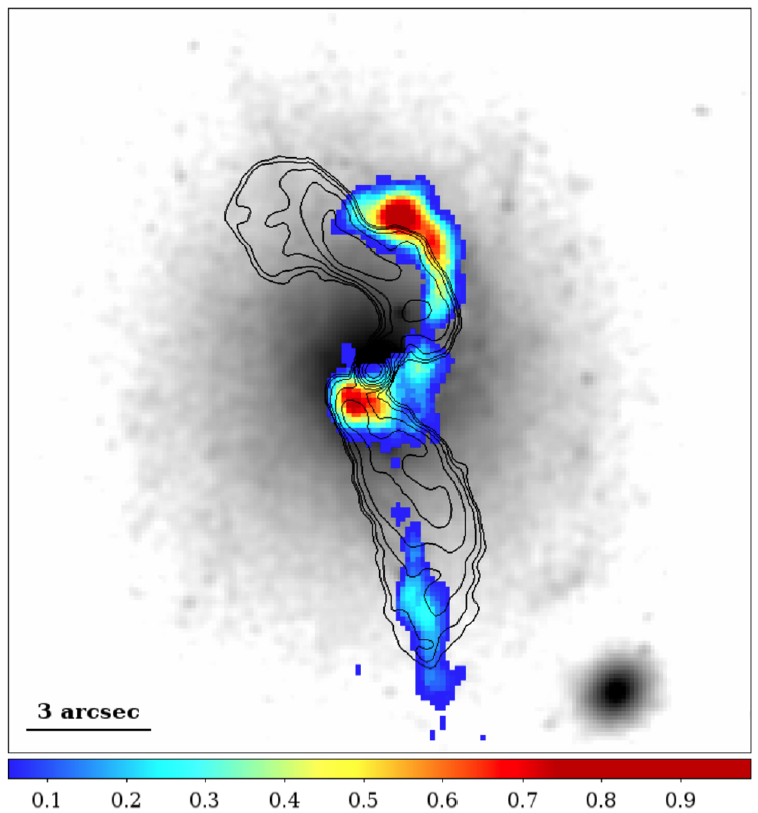}
   \includegraphics[width=0.4\textwidth]{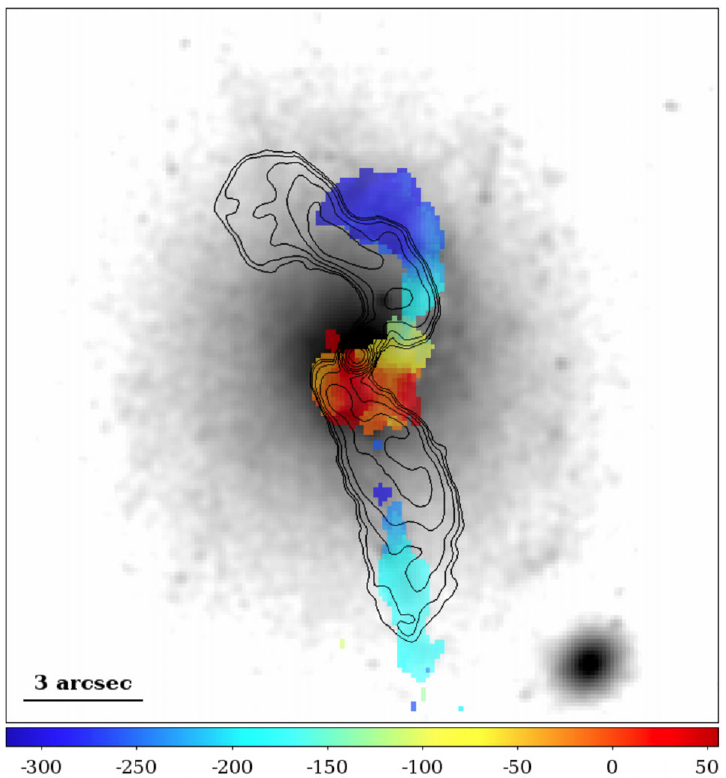}
   \caption{ {\bf Left:} ALMA CO(2-1) map of Abell 1795 (colour scale in Jy km/s), with
     the VLA 5GHz black contours superposed (from Breugel et al 1984),
     overlaid on the HST F702W image in gray-scale.
     {\bf Right:} CO(2-1) velocity map (colour scale from -300 to 50 km/s),
     overlaid on the HST image, and with the same VLA black contours.
From Rusell et al., 2017.}
   \label{fig1}
\end{center}
\end{figure}

\section{Molecular observations of cool core clusters}

A large number ($>$50) of cool core clusters have been observed now with ALMA,
NOEMA, or IRAM-30m, and large amounts of cold molecular gas have been detected, with
always the same features (e.g. Pulido et al. 2018).  There is a good correlation between
H$\alpha$ and CO emission, and molecular gas masses between  10$^9$ and 10$^{11}$ M$_\odot$
are well correlated with star formation. The global corresponding depletion time-scale
is just below 1 Gyr, somewhat smaller than the usual 2 Gyr value found in average
in nearby galaxies (Bigiel et al 2008).

In all resolved studies, where the molecular structures can be distinguished,
one strong feature is that the molecular gas (CO) and ionized gas (H$\alpha$) are co-spatial.
It is difficult to determine whether the filaments are inflowing or outflowing, but statistically
there must be both motions. Inflow has been proven in several cases through absorption (see below),
and this confirms that molecular gas is fueling the AGN. Outflow is also expected, due to the AGN feedback.
In the multiphase gas observed in the central 10-20kpc radii, thermal instabilities create turbulence,
clumps, which is necessary to provide more cooling, but certainly makes the kinematics more chaotic.
This has also the effect to mix the various phases (cold and hot) and homogeneize abundances,
explaining why the hot ICM is observed with relatively high metallicity.

One common characteristics of molecular components detected in cool-core clusters is
their low-velocity gradients in filaments, revealing that the filaments are not in free-fall,
and also the lack of relaxed structures, that could have settled in a rotating disk in the central BCG.
Already in Perseus, Salom\'e et al. (2008, 2011) have shown that the gas velocity in filaments is far from free-fall,
and that the gas around the center is not in rotation (or only in rare cases, like Hydra-A,  cf Hamer et al. 2014,
or very close to the center, in the central kpc).

Specific examples include the cool-core cluster Abell 1664, where ALMA has detected
an H$_2$ mass of 1.1 10$^{10}$M$_\odot$ in two velocity components (Russell et al. 2014):
one  -250  $<$ V $<$ 250km/s component
around the systemic velocity, and a high velocity one (HVS)
at -570km/s  (an outflow if in front of the BCG, or outflow otherwise). The structures are clympy
and complex, but without strong V-gradient.

In the cluster Abell 1835, ALMA has detected a molecular mass of 
M(H$_2$) = 5 10$^{10}$ M$_\odot$ within 10kpc of the BCG 
(McNamara et al. 2014). The velocity gradient is small, the profile narrow
(130km/s) meaning either an almost face-on disk, if in rotation, or inflowing gas to the center.
Ther is also a high velocity component, in which it is impossible to distinguish inflow from outflow,
some kpc from the center. However, the AGN is not powerful enough to match the energy budget,
and there is no bipolar counterpart. Besides the molecular gas is associated to an X-ray bubble, and could be
cold gas condensing from the hot ICM.

A very special case is the BCG in Abell 1795, which is sloshing
and oscillating in the central potential well of its cluster.
McDonald et al. (2009) have revealed a long trailing wake in H$\alpha$,
spanning $\sim$ 60 kpc in length, as a thin tail  behind the BCG,
and cospatial with an X-ray tail, with a bubble.
The dynamical time-scale of the oscillation is of the same order of the cooling
time, i.e. $\sim$ 300 Myr, and a corresponding molecular gas wake was detected
with IRAM interferometer (Salom\'e \& Combes 2004). ALMA results reveal
the molecular filaments with exquisite details (Russel et al. 2017).
As shown in Figure  \ref{fig1}, the molecular features are distributed around the
cavity sculpted by the radio jet in the ICM gas. The velocity map at the right of the Figure
shows no rotation, but small V-gradient, corresponding to slow inflow or outflow.

\begin{figure}[h]
\begin{center}
  \includegraphics[width=0.6\textwidth]{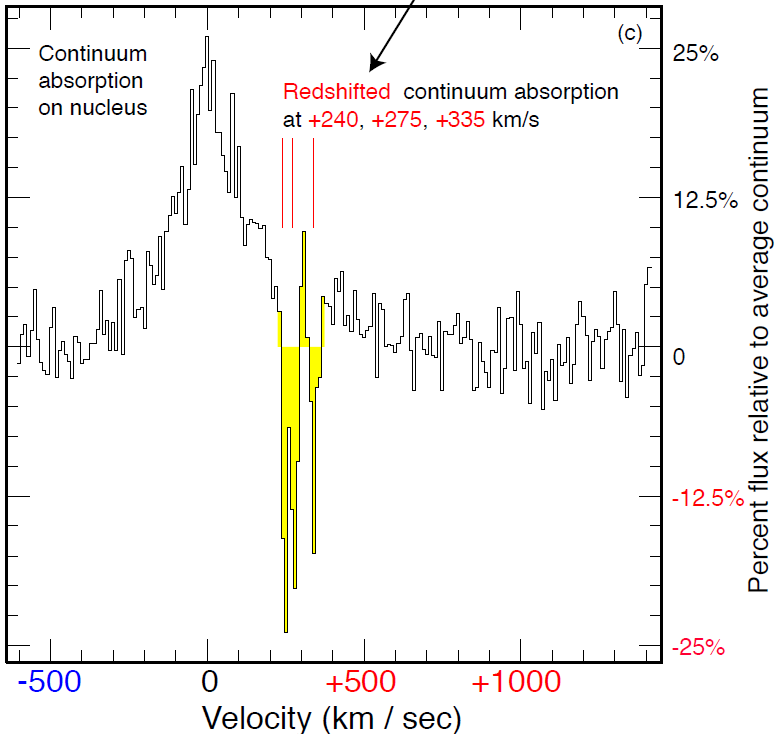}
  \caption{ ALMA CO(2-1) spectrum in front of the central continuum source
    in the galaxy cluster Abell 2597, a cool core cluster. The emission spectrum is seen
    at the systemic velocity, and then 3 absorbing components are redshifted (highlighted
    in yellow), indicating inflow of gas towards the nucleus.
From Tremblay et al., 2016}
   \label{fig2}
\end{center}
\end{figure}

In some cases, it is possible to distinguish between inflow or outflow,
when the molecular gas can be seen in absorption in front of the central AGN continuum source.
This is the case in the BCG of  Abell 2597, where  ALMA has revealed a CO(2-1) absorption  spectrum
(Tremblay et al. 2016, cf Figure \ref{fig2}). The CO emission is centered at the systemic velocity, but
the various absorption features are red-shifted only, suggesting inflowing
dense clouds, fueling the AGN.

Several cases of such molecular absorptions have been detected now (e.g. Edge et al. 2018, in prep)
and red-shifted ones are prevailing, although some 
are blue-shifted, when there is a signature of non-circular rotation, with a sharp edge
at terminal velocity and a broad wing. As for  Abell 2597, Tremblay et al. (2018) detect
a molecular mass of 3 10$^9$ M$_\odot$ within 30kpc, and probably inflow and
outflow at large distances, with also deviation of the jet by the molecular clouds.
Again there is no evidence for the gas to be in free-fall, but it shows some chaotic motions.
This might be explained through ram-pressure forces, if the molecular clouds reach
the velocity limit V$_l$. For an ICM density of $\rho_{ICM} \sim 0.1-0.2$ cm$^{-3}$ at  R$\sim$ 15-30kpc,
the corresponding column density $\rho_{ICM}$ R $\sim$  10$^{22}$ cm$^{-2}$, which is
similar to the column density in molecular clouds, at GMC scale and below.
The order of magnitude of the ram-pressure is then  $\rho_{ICM}$ V$_l^2$ A = M$_{cl}$ g =  M$_{cl}$ V$_l^2$/R,
where g is the local acceleration of the cluster.

\section{Expectations from theory and simulations}

The ubiquitous presence of X-ray cavities, bubbles filled with relativistic plasma,
ripples in the hot ICM, and multiple cold gas filaments directed straight to the cluster center,
or stretched around X-ray cavities and bubbles, suggest that the cold gas is condensing
from the hot ICM in the wakes of the buoyantly rising bubbles.
 Such a process of hot gas uplifting and then condensing in cold gas was simulated 
 by Revaz et al. (2008), producing a cycle of inflowing/outflowing multiphase gas,
 and McNamara et al. (2016) proposed a feedback mechanism in which
 the low entropy gas at the cluster center (with low t$_{cool}$), is entrained by the buoyantly rising bubbles,
 inflated by the AGN radio jets, to an altitude of $\sim$ 10 kpc, where
 the cooling time then becomes lower than the infall time t$_I$, which can be a few
 times longer than  t$_{ff}$. The free-fall time indeed becomes much longer at
 high altitude R, being proportional to (R/g)$^{1/2}$, where g is the local
 gravitational acceleration. In the adiabatic uplift, t$_{cool}$ on the contrary
 varies very little.

 The AGN feedback has been recognized as the main moderator
 of the huge cooling flows that were previously expected (e.g. Fabian 2012),
 and now thermal instabilities (TI)  (McCourt et al.  2012) are shown to be
 key in allowing the gas to cool far from the center, at R=10-20kpc, as soon as
 t$_{cool}$/t$_{ff} \sim$ 10. The radio jets inflate bubbles, which  create inhomogeneities
 and trigger further cooling. This has been dubbed
 Chaotic Cold Accretion  (CCA) by Gaspari et al. (2011, 2012).

 This phenomon is perfectly compatible with the kinematics of the gas
 observed in cool core clusters. For instance, the 
 Hitomi satellite in Perseus has shown that the gas velocity dispersion
 $\sigma_{gas}\sim$ 160 km/s, at 30-60 kpc  (in the Fe XXV-Fe XXVI plasma, Lau et al. 2017).
 This implies a rather quiescent AGN feedback, in spite of cavities, and inhomogeneities.
This gentle kinetic feedback corresponds more to the CCA
than to violent galaxy cluster mergers. Cool core clusters appear relaxed since 4 Gyr.

Observations and simulations tend to reveal the existence of
a floor in the value of the ratio  t$_{cool}$ /t$_{ff}$, from a typical radius
of $\sim$ 10kpc down to the cluster center. While the outer atmosphere,
with  t$_{cool}$ /t$_{ff} >>  10$ is quite stable, the thermal instability
and cooling occurs at a few 10kpc. Cooling is most efficient in the presence
of TI because of the strong dependence of the cooling rate on density.

Sharma et al (2012) have shown how local thermal instability (TI)
occurs whenever the TI time scale becomes lower than 10 times the free-fall time.
Self-regulation ensures a lower-limit  in the ratio t$_{TI}$ /t$_{ff}$, and can explain the
floor. Voit et al. (2015) have called precipitation the combined process of
TI and buoyancy, by analogy with the precipitation of rain drops.
This phenomenon is quite similar to what Gaspari et al. (2011, 2012) call
chaotic cold accretion (CCA) in their simulations. This kind of accretion
can be much more efficient than the previous simple model of a Bondi
accretion from the hot ICM, and corresponds much better to observations.
Gaspari \& Sadowski (2017) have tried to unify the widely different scales occuring
in groups and clusters, under the self-regulation of AGN feedback, from the
ultra-fast outflow (UFO) at scale smaller than a parsec, to kpc scales where
thermal instabilities occur, and the outflowing gas comes back, even if launched
with the escape velocity.

In a recent generalizing paper, Voit et al. (2017) describe in detail the history of
ideas in thermal instabilities, and the new important effect brought by the combination with
the buoyancy.  The latter, intimately linked to AGN feedback,
is essential to uplift the gas in the center of cluster, but can
be replaced by the sloshing of the BCG at the bottom of the cluster potential well, or
by mergers. 

That energetic feedback is able to  trigger cold gas accretion from a hot atmosphere,
is also seen in many other circumstances, in haloes of various masses. There is an
analogy with normal spiral galaxies, where the feedback is now supernovae or stellar
feedback, inducing a fountain effect.  This has been described for the halo of the Milky
Way or galaxies such as NGC 891, where cold accretion of HI gas has been observed
(e.g. Fraternali \& Binney 2008, Marinacci et al. 2011).
Simulations have shown how some gas from the hot corona could be made cooling,
and fueling the gas disk. The cooling time is very large, and without the supernovae feedback,
the corona would be stable. Also it is difficult even in flat-entropy coronae to trigger
thermal instabilities, in the presence of thermal conduction and buoyancy. However, the
fountain effect helps the hot corona to cool: as the fountain gas is moving in the corona,
it is stripped off and mixed with the hot coronal gas, and since the metallicity is then high
enough, it can radiatively cool. The fountain gas drags a cooling tail in its wake, so that
a net fueling of the disk occurs at a few M$_\odot$/yr, about equal to the global star formation
rate of the galaxy (Marasco et al. 2013), ensuring self-regulation.
This mechanism explains  the presence of multiphase gas in the atmospĥere of
galaxies.

\begin{figure}[h]
\begin{center}
  \includegraphics[width=0.6\textwidth]{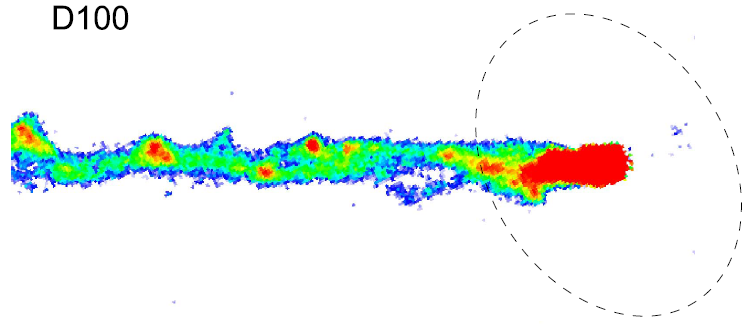}
   \includegraphics[width=0.3\textwidth]{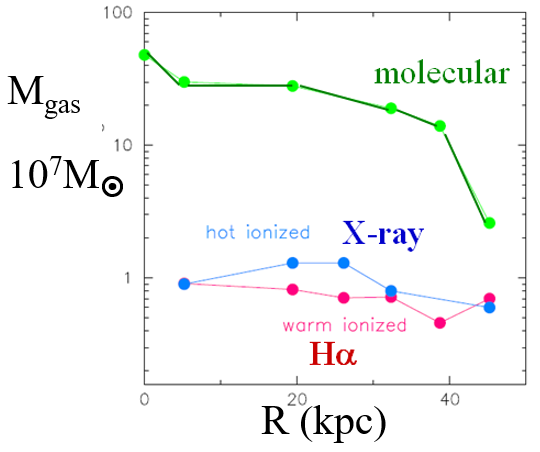}
   \caption{ {\bf Left:} H$\alpha$ image of the D100 ram-pressure tail
     in Coma, from Yagi et al. 2007. The silhouette of the spiral galaxy is
     indicated with a dashed ellipse. The fraction of the tail shown is 20kpc long.
     {\bf Right:} Masses of the various gas components, molecular gas, hot X-ray gas and
 warm H$\alpha$ gas, in 12'' beams (the IRAM CO(2-1) beam)
     along the tail of D100, until 50kpc from the center.
From Jachym et al., 2017}
   \label{fig3}
\end{center}
\end{figure}

\section{ Molecular gas in X-ray groups}

Cold gas condensing from the hot X-ray atmospheres is also expected
at smaller scale in galaxy groups. One typical example is NGC 5044, at the center
of a small group, characterized by multiple and radial H$\alpha$ filaments, co-spatial with
X-ray structures mapped with Chandra. The molecular gas was detected by ALMA, in a series
of giant molecular associations (GMA), with masses between 
3 10$^5$ and 10$^7$ M$_\odot$, and velocity wisths 10-50km/s. There is
no rotating disk, but clumps also seen in absorption, only red-shifted (David et al. 2014).

Comparable features have been detected in two other groups, NGC 4638 and NGC 5846,
by Temi et al (2018). The gas clouds are aligned on H$\alpha$ features, and delineated by dust lanes.
The origin of the cold gas is cooling from the hot medium (which is dust-enhanced due to previous mixing).
The clumps or GMA are not bound, their Virial ratio  being much  larger than 1  (20-200).
These observations are compatible with the same CCA mechanism.

\section{Origin of the molecules}

In these hostile and hot environments, how can dense and cold molecular clouds
survive, in the presence of strong shocks due to the AGN feedback and radio jets?
Paradoxically, although the energetic incident radiation, or strong jets can
destroy the clouds and make them more diffuse, they not only increase
the surface of the clouds, but produce Kelvin-Helmholtz instabilities,
which trigger the  re-formation of molecules downstream
(Hopkins \& Elvis 2010).
Some numerical simulations found that destroyed clouds do not re-condense,
and they do not offer enough surface to be entrained (Ferrara \& Scannapieco 2016).
Dust is destroyed  by sputtering, and dust is a necessary catalyzer for H$_2$ molecule formation.
New simulations, taking into account chemistry, H$_2$ formation, and all the dust cycle
have shown however that molecules can reform in AGN outflows, with the
right order of magnitude (Richings \& Faucher-Gigu\`ere 2018).

\begin{figure}[b]
\begin{center}
 \includegraphics[width=0.8\textwidth]{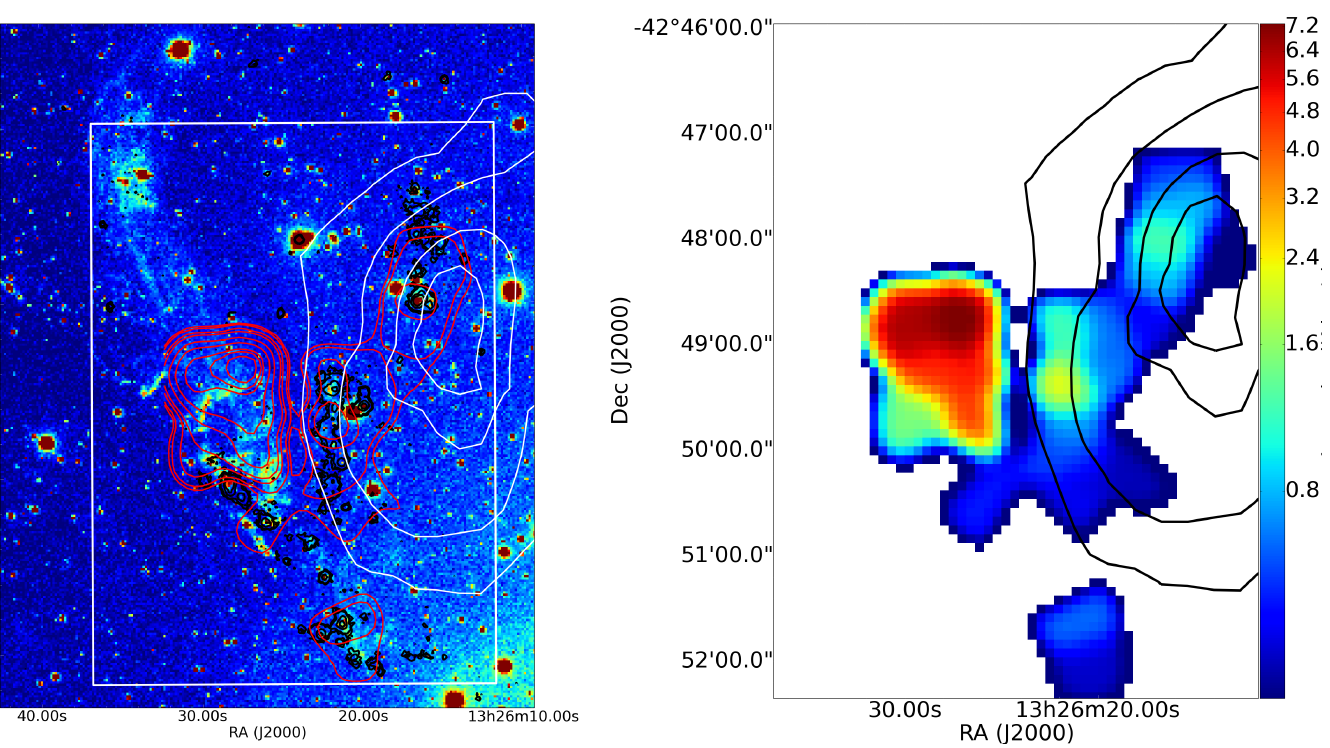}
 \caption{ {\bf Left:} H$\alpha$
emission of the northern region of Centaurus A with CTIO. The HI
emission (VLA; white contours), the CO(2-1)emission
(APEX; red contours) and the FUV emission (GALEX; black contours) are overlaid.
{\bf Right:} intensity map of the CO(2-1) emission from APEX
in K km/s, with the HI emission overlaid in black contours. The white box on the left shows
the region of the right panel. From Salom\'e et al. 2016.
}
   \label{fig4}
\end{center}
\end{figure}

\section{Molecular filaments in ram-pressure stripped tails}

Other molecular filaments, co-spatial to H$\alpha$ filaments can be
seen in galaxy clusters, this time due to the cluster environment
perturbations (tidal and ram-pressure tails).
A giant H$\alpha$  tail, larger than 100 kpc, has been  detected in Virgo
by Kenney et al. (2008). This tail begins at NGC 4438, and turns around
M86, and at both extremities, CO emission has been detected
(Dasyra et al. 2012). This shows that  H$_2$ gas can survive more than 100 Myr
in hostile environments, at temperature of 10$^7$K.
Clouds of mass M(H$_2$) =7 10$^6$M$_\odot$ are seen 10kpc NE of M86.
These have been formed in situ or come from the NGC 4438  tail. 

In rich clusters, the ram-pressure stripping is even more violent,
as demonstrated by the ESO137-001 tail in the Norma cluster
(Jachym et al. 2014). Along 80kpc, a hot X-ray tail is co-spatial to warm H$\alpha$
gas and molecular gas, emitting in CO. The tail is de-doubled, certainly due to the two sides
of the rotating disk. In Coma, the galaxy D100 reveals an even thinner ram-pressure tail,
which is the second stage of stripping, when the outer parts have been swept out already
(Jachym et al. 2017). Figure \ref{fig3} shows the straight H$\alpha$ filament (left)
and how the mass of the filament is dominated by molecular gas (right).

AGN feedback due to radio jets can also be positive, and trigger star formation
in the hot atmosphere of a giant elliptical galaxy. In  Centaurus A, a long filament of young
star formation aligned along the radio jet is known since a long time. ALMA and APEX
have detected large quantities of molecular gas aligned on this H$\alpha$ filament,
just when the radio jet encountered an HI shell, cf Figure \ref{fig4}.. On the jet path, all the atomic gas has
been transformed in molecules (Salom\'e et al. 2016).

\section{Summary}

Molecular filaments have now been detected in many cool-core clusters,
with common characteristics: they form in the wake of X-ray bubbles, inflated
by the AGN radio jets, they are not in free fall, but reveal small velocity
gradients. Observations are compatible with their formation by precipitation of cold gas,
or chaotic cold accretion (CCA) in a thermally unstable hot medium, as soon as
t$_{cool}$/t$_{ff} \sim$ 10. When the entropy profile is relatively shallow in the center
of clusters, some low entropy gas can be adiabatially uplifted until this ratio is
sufficiently small, and the gas can cool, in a multiphase medium.

This cold gas fuels the central AGN, which can then  self-regulate
the heating and cooling to the balance observed. Thermal instabilities,
combined with buoyancy, and AGN feedback can create the conditions of CCA
up to 10-20 kpc from the cluster center. The gas in filaments is slowed down from
free-fall through ram-pressure forces.

When the BCG is in oscillating motion in the bottom of the potential
well, a cooling wake is formed, with X-ray, H$\alpha$ and CO emission co-spatial
in the wake.
Molecules can reform easily in the filaments

Molecular filaments are also present in tidal and ram-pressure tails.
The cooling of hot coronae can also be triggered by supernovae feedback:
fountain gas ejected above the plane of spiral galaxies is then mixing with
hot gas, triggering cooling, with a net balance of fueling star formation in the disk.
AGN feedback can also be positive, as shown by very clear cases of jet-induced star formation.

\end{document}